\documentclass{mn2e}
\usepackage[dvips]{graphicx}
\usepackage{amssymb}
\usepackage{times}

\newcommand{\source}{4U~1820--303}
\newcommand{\msun}{{\rm M}_{\sun}}
\newcommand{\rsun}{{\rm R}_{\sun}}
\newcommand{\xte}{{\it RXTE}}

\title[The triple nature of \source]
{The superorbital variability and triple nature of the X-ray source \source}

\author[A. A. Zdziarski, L. Wen, M. Gierli\'nski]
{Andrzej A. Zdziarski,$^1$\thanks{E-mail:
aaz@camk.edu.pl} Linqing Wen,$^2$\thanks{E-mail: lwen@aei.mpg.de} and Marek Gierli\'nski$^{3,4}$\thanks{E-mail: Marek.Gierlinski@durham.ac.uk}\\
$^1$Centrum Astronomiczne im.\ M. Kopernika, Bartycka 18, 00-716 Warszawa, Poland\\
$^2$Max-Planck-Institut f\"ur Gravitationsphysik, Albert-Einstein-Institut, Am M\"uhlenberg 1, D-14476 Golm, Germany\\
$^3$Department of Physics, University of Durham, Durham DH1~3LE, UK\\
$^4$Obserwatorium Astronomiczne Uniwersytetu Jagiello\'nskiego, Orla 171, 30-244 Krak{\'o}w, Poland
}

\date{Accepted 2007 February 28. Received 2007 January 16}

\pagerange{\pageref{firstpage}--\pageref{lastpage}}
\pubyear{2007}

\begin{document}

\maketitle

\label{firstpage}

\begin{abstract}  
We perform a comprehensive analysis of the superorbital modulation in the ultracompact X-ray source \source, consisting of a white dwarf accreting onto a neutron star. Based on \xte\/ data, we measure the fractional amplitude of the source superorbital variability (with a $\sim$170-d quasi-period) in the folded and averaged light curves, and find it to be by a factor of $\sim$2. As proposed before, the superorbital variability can be explained by oscillations of the binary eccentricity. We now present detailed calculations of the eccentricity-dependent flow through the inner Lagrangian point, and find a maximum of the eccentricity of $\simeq 0.004$ is sufficient to explain the observed fractional amplitude. We then study hierarchical triple models yielding the required quasi-periodic eccentricity oscillations through the Kozai process. We find the resulting theoretical light curves to match well the observed ones. We constrain the ratio of the semimajor axes of the outer and inner systems, the component masses, and the inclination angle between the inner and outer orbits. Last but not least, we discover a remarkable and puzzling synchronization between the observed period of the superorbital variability (equal to the period of the eccentricity oscillations in our model) and the period of the general-relativistic periastron precession of the binary. 
\end{abstract}
\begin{keywords}
accretion, accretion discs -- binaries: general -- globular clusters: individual: NGC 6624 -- stars: individual: 4U~1820--303 --  X-rays: binaries -- X-rays: stars.
\end{keywords}

\section{INTRODUCTION}
\label{intro}

\source\ is one of the most remarkable low-mass X-ray binaries. This ultracompact binary consists of an $M_2=(0.06$--$0.08)\msun$ He white dwarf secondary (Rappaport et al.\ 1987, hereafter R87) accreting via Roche-lobe overflow onto a neutron star (of the mass $M_1$). Its binary period is as short as $P_1\simeq 685$ s (Stella, Priedhorsky \& White 1987; Anderson et al.\ 1997). A likely scenario for the formation of \source\ appears to be a direct collision of a neutron star and a giant (Verbunt 1987; Ivanova et al.\ 2005) in its parent globular cluster NGC 6624. This resulted in a binary consisting of the neutron star and the stripped giant core, which then cooled down to the present relatively degenerate state (R87). The secondary has to have a very low H abundance in order to fit its Roche lobe (R87), and the occurence of type-I X-ray bursts implies that it cannot be made of elements heavier than He. Interestingly, \source\ was the first X-ray burster identified with a known X-ray source (Grindlay et al.\ 1976). The most likely distance to the source appears to be $7.6\pm 0.4$ kpc (Kuulkers et al.\ 2003), as discussed in a companion paper (Zdziarski et al.\ 2007, hereafter Z07).

The accretion occurs as a consequence of the loss of the angular momentum of the binary via emission of gravitational radiation (Paczy\'nski 1967), as proposed for this system by Stella et al.\ (1987) and Verbunt (1987), and calculated in detail by R87. The implied present mass-loss rate from the secondary (R87; see Section \ref{flow} below) is completely compatible with that corresponding to the average isotropic bolometric luminosity measured from the {\it Rossi X-ray Timing Explorer\/} (\xte) data by Z07. This provides a strong support for this accretion model. The present phase of the high mass transfer started only $\sim\! 10^6$ yrs ago and may be sustained only for $\sim\! 3\times 10^6$ yrs (R87).

The 685-s period was discovered in X-rays as a modulation with a $\sim$2--3 percent peak-to-peak amplitude (Stella et al.\ 1987). The period is very stable, with a low $\dot P_1/P_1=(-3.5\pm 1.5)\times 10^{-8}$ yr$^{-1}$ (Chou \& Grindlay 2001, hereafter CG01), which makes it certain it is due to the binary motion. The modulation was proposed to be due to the obscuration of the X-ray source by a structure at the edge of the accretion disk. A stronger modulation in the UV was predicted by Arons \& King (1997), and subsequently discovered by Anderson et al.\ (1997), with $P_1= 687.6\pm 2.4$ s and a 16 per cent peak-to-peak amplitude. 

The most unusual feature of \source\ appears to be the luminosity variation by factor of $\ga$2 at a (superorbital) period of $P_3\simeq 170$ d (Priedhorsky \& Terrell 1984; Smale \& Lochner 1992; CG01; {\v S}imon 2003; Wen et al.\ 2006, hereafter W06). CG01 found the modulation is stable with $P_3=171.0\pm 0.3$ d and $\vert \dot P_3/P_3\vert <2.2\times 10^{-4}$ yr$^{-1}$ based on data from 1969 to 2000. W06 found $P_3=172\pm 1$ d based on 8.5 yrs of \xte\/ All Sky Monitor (ASM) data. The fact that X-ray bursts take place only at the flux minima (Cornelisse et al.\ 2003, Z07) proves that the observed variability is due to intrinsic luminosity/accretion rate changes and not due to, e.g., obscuration or changes of the projected area of the source due to precession. This is further supported by strong correlations between the observed flux and the source spectral state, varying with the flux in a way typical of atoll-type neutron-star binaries (Bloser et al.\ 2000; Gladstone, Done \& Gierli{\'n}ski 2007), and between the frequency of kHz QPOs observed from the source and the flux (Zhang et al.\ 1998; van der Klis 2000). The source was classified as an atoll by Hasinger \& van der Klis (1989). 

The intrinsic, accretion-rate related, character of the long-term periodic flux changes in \source\ is unique among all sources showing superorbital variability. In other cases, such changes of the observed flux appear compatible with being caused by accretion disc and/or jet precession, which either results in variable obscuration of emitted X-rays as in Her X-1 (Katz 1973), or changes the viewing angle of the presumed anisotropic emitter, as in SS 433 (Katz 1980) or Cyg X-1 (e.g., Lachowicz et al.\ 2006), or both. However, that precession keeps the inclination angle of the disc with respect to the orbital plane constant, and thus it cannot change the accretion rate (or the luminosity). Also, the ratio between the superorbital and orbital periods is $\simeq 2.2\times 10^4$, which is much higher than that possible to obtain from any kind of disc precession at the mass ratio of the system (e.g., Larwood 1998; Wijers \& Pringle 1999). This appears to rule out also models which would attempt to explain the variable accretion rate by a varying disc inclination angle (with respect to the orbital plane). 

In order to explain the long-term periodic variability of the accretion rate, CG01 proposed a hierarchical triple stellar model, in which a distant tertiary exerts tidal forces on the inner binary. This results in a cyclic exchange of the angular momentum between the inner system and the tertiary (Kozai 1962; Lidov \& Ziglin 1976; Mazeh \& Shaham 1979; Bailyn \& Grindlay 1987; Ford, Kozinsky \& Rasio 2000, hereafter F00; Blaes, Lee \& Socrates 2002; Wen 2003, hereafter W03). The period of the resulting evolution of the parameters of the system is $\sim\! P_2^2/P_1$, where $P_2$ is the orbital period of the tertiary, implying $P_2\sim 1$ d. The variable system parameter relevant here is the eccentricity, $e_1$, of the inner system, which causes changes of the distance between the inner Lagrange point, $L_1$, and the center of mass of the secondary. This in turn changes the rate of the flow through $L_1$ and the accretion rate. The mass of the tertiary has been constrained by CG01 as $M_3\la 0.5\msun$ based on the lack of its optical detection. We note that this constraint assumes the third body is a main-sequence star; if it is a white dwarf or a neutron star, $M_3\la 1.4\msun$. (Still, the third star is most likely on the main sequence, which we assume hereafter.) However, no specific calculations of the proposed triple system, e.g., of the maximum $e_1$ from modelling the variable rate of the flow through the $L_1$ point, or of the system parameters from modelling gravitational evolution of the system, have been done yet. 

Here, we first analyse the \xte\/ ASM monitoring data and the Proportional Counter Array (PCA) scanning data, and use them to quantify the source X-ray variability. We then present calculations on the dependence of the accretion rate through $L_1$ on the eccentricity, which yields the maximum $e_1$ required to reproduce the observed amplitude of the superorbital variability. Then, we model evolution of hierarchical triple stellar systems, and reproduce the $\sim$170-d period and the maximum eccentricity implied by the $L_1$-flow model. Our results put constraints on the masses of the system component, the inclination between the inner and outer orbits, and the ratio between its semimajor axes. We also report a discovery of a remarkable synchronization between the superorbital period and the period of the relativistic periastron precession of the binary. 

\section{The data}
\label{data}

\begin{figure*}
\centerline{\includegraphics[width=115mm]{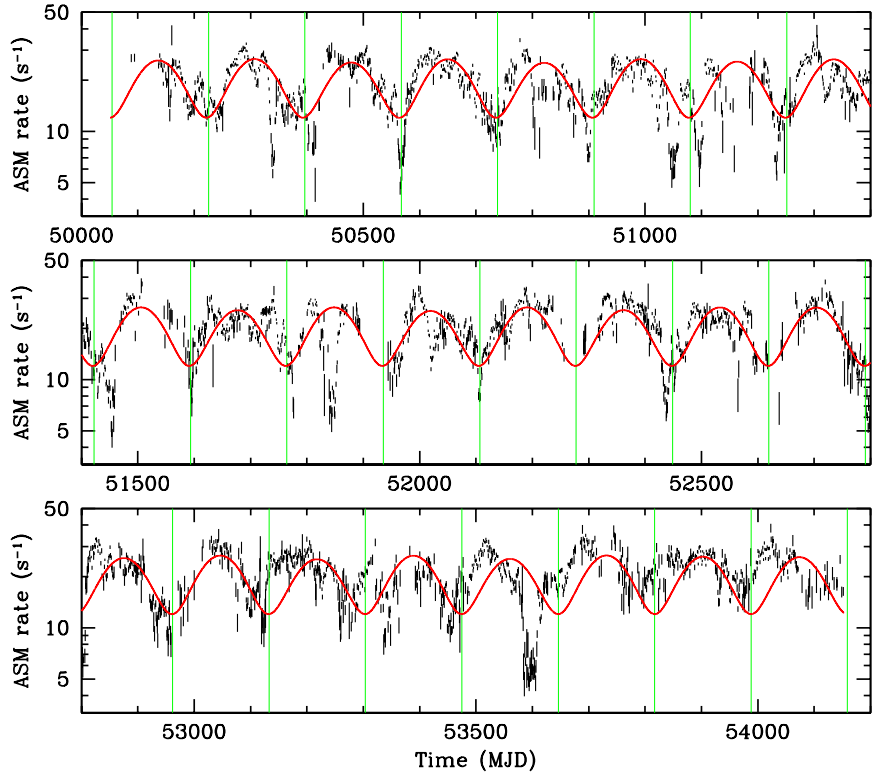}}
\caption{The \xte/ASM light curve based on 1-day average measurements. The vertical lines show the minima of the superorbital cycle according to the ephemeris of CG01. Note departures of the ASM minima from the predicted dates up to $\sim$50 d in some cases, as well as secondary minima (see also {\v S}imon 2003). The solid curve presents our theoretical model of Sections \ref{flow}--\ref{triple}.
\label{asm} }
\end{figure*}

Fig.\ \ref{asm} shows the long-term light curve of \source\ from the \xte\/ ASM (1996 January 1--2007 February 20). For the clarity of display, we have averaged some adjacent 1-day measurements in order to achieve the minimum significance of $3\sigma$ of the plotted count rate. We clearly see the cyclic variations of the count rate on the $\sim$170-d quasi-period with a large relative amplitude. The vertical lines show the predicted minima according to the ephemeris of eq.\ (9) of CG01. Using the Lomb-Scargle periodogram (Lomb 1976; Scargle 1982), we find that the present (daily-averaged) ASM data yield $P_3=170.6 \pm 0.3$ d, compatible with the periods of CG01 and of W06. 

\begin{figure}
\centerline{\includegraphics[width=84mm]{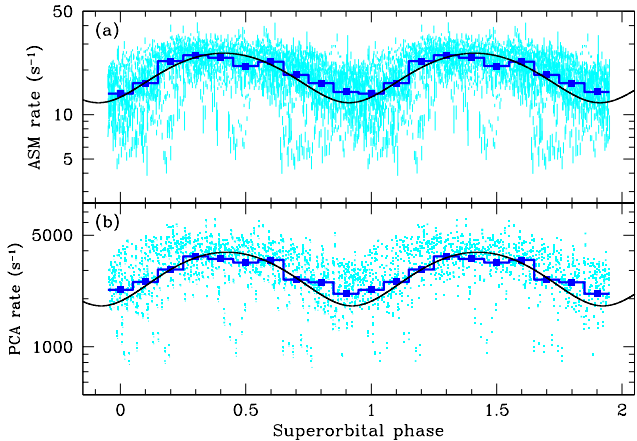}}
\caption{(a) The ASM light curve of Fig.\ \ref{asm} folded on the superorbital ephemeris of CG01. (b) The folded light curve for the PCA Galactic bulge scan data (see Section \ref{data}). On each panel, the histogram shows the light curve averaged over 10 phase bins, and the solid curve shows the theoretical model of Sections \ref{flow}--\ref{triple} (based on the second cycle of Fig.\ \ref{f_triple}a) fitted to the averaged data. 
\label{phase} }
\end{figure}

Fig.\ \ref{phase}(a) shows the ASM light curve folded on the ephemeris of CG01, as well as the folded light curve averaged over 10 phase bins. We see that while the daily measurements span a factor $\ga$10, the averaged light curve varies over the superorbital period spanning a factor of $\simeq$2. 

We also use the available Galactic bulge scan data\footnote{lheawww.gsfc.nasa.gov/users/craigm/galscan/html/4U\_1820-30.html} from the \xte\/ PCA detector for the time interval of 1999 February 5--2006 October 30. Fig.\ \ref{phase}(b) shows the count rate from those measurements folded, and folded over the superorbital ephemeris. We also show the count rate folded and averaged over 10 phase bins. Similarly to the ASM data, we see that while the individual measurements of the count rate span a factor $\simeq$10, the averaged light curve varies over the superorbital period spanning a factor $\simeq$2. We stress, however, that the variability is not strictly periodic, and $P_3$ represents only a quasi-period. Therefore, folding and averaging over a single period value results in some suppression of the actual average superorbital variability. With that caveat, we attribute the changes of the average flux within the factor $\simeq$2 to a relative stable quasi-periodic process (Sections \ref{flow}--\ref{triple}), and the remaining variability to some aperiodic processes, in particular those operating in other atoll sources. We also assume that the ASM and PCA count rate variability reflects that of the accretion rate, and assume hereafter its superorbital variability consists of variations within (0.7--$1.4)\langle -\dot M_2\rangle$.

Z07 have, in addition, analyzed available PCA and High Energy X-Ray Timing Experiment (HEXTE) pointed observations of \source. Using a physically-motivated spectral model, they have estimated the bolometric flux from the source for each observation. The resulting light curve, when folded over the ephemeris, is very similar to those of Fig.\ \ref{phase}. In particular, the fractional variability of the folded/averaged light curve is also found to be $\sim$2. Since the bolometric flux is highly likely to be proportional to the rate of accretion onto a neutron star, that result confirms that the superorbital range of $-\dot M_2$ is indeed within a factor of $\sim$2, as adopted above.

\section{The mass flow through the inner Lagrangian point}
\label{flow}

Here, we calculate the dependence of the rate of the Roche lobe overflow (assumed to equal the accretion rate) on the eccentricity. For that, we follow classical results on the Roche flow, expressing $\dot M_2$ as the product of the density and the sound speed at the inner Lagrangian point, $L_1$, times an effective area of the flow. The position of the $L_1$ varies with the orbital phase in an eccentric orbit, yielding a varying $\dot M_2$. This yields the orbit-averaged $\dot M_2$ as a function of the eccentricity, $e_1$. This in turn allows us to determine the maximum eccentricity, $e_{\rm max}$, required to enhance $\dot M_2(e)$ by the factor of two inferred above from the varying $F_{\rm bol}$, assuming that the minima of the superorbital cycle correspond to the $e_1$ minimum of $e_0\simeq 0$. 

We first briefly estimate the parameters of the inner binary. We assume that the white dwarf fills its Roche lobe. Then, we combine the volume-averaged Roche-lobe radius for $M_2/M_{12}\la 0.8$ of Paczy\'nski (1971), \begin{equation}
{R_2\over a_1}= {2\over 3^{4/3}} \left(M_2\over M_{12} \right)^{1/3},
\label{roche}
\end{equation}
with the Kepler law, 
\begin{equation}
a_1^3=GM_{12} {P_1^2\over 4\upi^2},
\label{kepler}
\end{equation}
where $G$ is the gravitational constant, $a_1$ is the semimajor axis, and $M_{12}=M_1+M_2$. This yields the relation,
\begin{equation}
P_1={9\upi\over 2^{1/2}} {R_2^{3/2}\over (G M_2)^{1/2}}.
\label{m_r}
\end{equation}
We then use the mass-radius relation for cold, low-mass stars of Zapolsky \& Salpeter (1969; as fitted in R87), and assume pure He and the radius equal that of the fully degenerate configuration times a factor, $f_{\rm d}$. For $f_{\rm d}=1.1\pm 0.1$ (R87), $M_2\simeq 0.067_{-0.010}^{+0.010}\msun$ and $R_2\simeq 2.19^{+0.10}_{-0.11}\times 10^9$ cm, where the lower and upper limits correspond to $f_{\rm d}=1$ and 1.2, respectively. Hereafter, we assume $f_{\rm d}=1.1$, for which the above $M_2$ and $R_2$ correspond to an He model of Deloye \& Bildsten (2003) (who has calculated the mass-radius relation for white dwarfs of arbitrary degeneracy\footnote{We note that for a cold, fully degenerate, white dwarf with the parameters relevant to \source, the model of Deloye \& Bildsten (2003) gives somewhat lower radii than that of Zapolsky \& Salpeter (1969).}) with the core temperature of $\simeq 10^7$ K and the core density of $\simeq 10^{4.2}$ g cm$^{-3}$. We note that for a star approximated by an $n=3/2$ polytrope (Chandrasekhar 1939), we have
\begin{equation}
R_2\simeq 0.0128 (1+X)^{5/3} \left(M_2\over \msun\right)^{-1/3} \rsun, 
\label{rwd}
\end{equation}
(where $X$ is the H mass fraction), which expression combined with equation (\ref{m_r}) at $X=0$ yields $M_2\simeq 0.067\msun$ and $R_2\simeq 2.19\times 10^9$ cm, the values identical to those above for $f_{\rm d}=1.1$. Then $a_1\simeq 1.29\times 10^{10} (M_{12}/1.35\msun)^{1/3}$ cm, where we used $M_{12}=1.35\msun$ corresponding to the presence of the resonance implied by equation (\ref{pgr0}) below, which then yields $M_1\simeq 1.28\msun$. 

For those parameters, and assuming negligible effects of a possible outflow, eq.\ (17) of R87 yields the average mass transfer rate of $\langle-\dot M_2\rangle\simeq 3.5^{+1.1}_{-0.9}\times 10^{17}$ g $s^{-1}$ (where the lower and upper limits correspond to $f_{\rm d}=1$, 1.2, respectively). This rate is fully compatible with the accretion rate, $\simeq (3.3\pm 0.1)\times 10^{17}$ g $s^{-1}$ (where the error is statistical only), corresponding to the average bolometric flux measured using the pointed \xte\/ PCA/HEXTE observations, $\langle F_{\rm bol}\rangle \simeq (8.7\pm 0.2)\times 10^{-9}$ erg cm$^{-2}$ s$^{-1}$ (Z07) at $D=7.6$ kpc and an accretion efficiency of 0.2. Using $M_1=1.4\msun$ increases the above theoretical rate by $\sim 0.2\times 10^{17}$ g $s^{-1}$. In our numerical estimates below, we use $M_1=1.28\msun$, $M_2= 0.07\msun$, $R_2= 2.2\times 10^9$ cm, and the corresponding $\langle-\dot M_2\rangle = 4\times 10^{17}$ g $s^{-1}$. 

The distance, $R_0$, between the center of the white dwarf and the $L_1$ point at $e_1=0$, is given by the solution of 
\begin{equation}
(1-x)^{-2} -(1-x)={M_2\over M_1}(x^{-2}-x),
\end{equation}
where $x\equiv R_0/a_1$, which yields $R_0\simeq 0.237 a_1\simeq 3.05\times 10^9$ cm. In an elliptical orbit, the separation between the stars at a given orbital angle, $\phi$ (measured with respect to the periastron), is given by,
\begin{equation}
s_1={a_1 (1-e_1^2)\over 1+e_1 \cos \phi},
\label{separation}
\end{equation}
and the angular velocity is
\begin{equation}
{{\rm d\phi}\over {\rm d}t}={2\upi\over P_1} {(1+e_1 \cos \phi)^2 \over
(1-e_1^2)^{3/2} }.
\label{angular}
\end{equation}
At $e_1\ll 1$, the $L_1$ distance is proportional to the separation, i.e., it equals $R_0+d'$, where
\begin{equation}
{d'\over R_0}\simeq - e_1 {e_1+\cos \phi\over 1+e_1 \cos \phi},
\label{e_depth}
\end{equation}
which changes between $-e_1$ and $e_1$. (See Brown \& Boyle 1984 for an expression valid at large $e_1$.)

The rate of the Roche lobe overflow can be generally written as (e.g., Savonije 1983),
\begin{equation}
-\dot M_2= {\cal A} \rho c_{\rm s},
\label{mdot}
\end{equation}
where ${\cal A}$ is an effective area of the flow, $\rho$ and $c_{\rm s}$ are the mass density and the isothermal sound speed, $c_{\rm s}=(kT/\mu m_{\rm H})^{1/2}$, respectively, at $L_1$, and $\mu$ is the mean molecular weight. Thus, $\dot M_2$ depends on the depth of the $L_1$ point within the donor star, $d$. It can be shown by expanding the effective gravitational potential around $L_1$ that
\begin{equation}
{\cal A}\simeq {P_1^2\over 2\upi} {GM_2\over R_0^2} d,
\label{area}
\end{equation}
(Savonije 1983). (Note that distinguishing here between the radius of the $L_1$ point, $R_0$, and the volume-averaged radius of the star filling the Roche lobe, $R_2$, would require much more complex treatment of the flow than that adopted here.) Paczy\'nski \& Sienkiewicz (1972) find that in the polytropic case $-\dot M_2\propto d^{1.5+n}$, where $n$ is the polytropic index [which can be shown to follow from equations (\ref{mdot}--\ref{area})]. If we neglect for a while the illumination of the white dwarf by the X-ray source, the relevant index would be $n=3.25$ of the (non-degenerate) surface layers of the white dwarf (Schwarzchild 1958). Then, the orbital-angle dependent accretion rate in an elliptical orbit can be written as,
\begin{equation}
{-\dot M_2(\phi, e_1)\over \dot M_0} \simeq  \left[\max(d_0-d',0)\over d_0\right]^{4.75},
\label{mdot_cold}
\end{equation}
where $d_0$ and $\dot M_0$ are the depth of the $L_1$ point and the accretion rate, respectively, at $e_1=0$ (and $d=d_0-d'$). Hereafter we assume that the minimum of the average light curve (see Fig.\ \ref{phase}) corresponds to $e_1= 0$, i.e., $\dot M_0 =0.7\langle -\dot M_2\rangle$ (see Section \ref{data}). 

\begin{figure}
\centerline{\includegraphics[width=84mm]{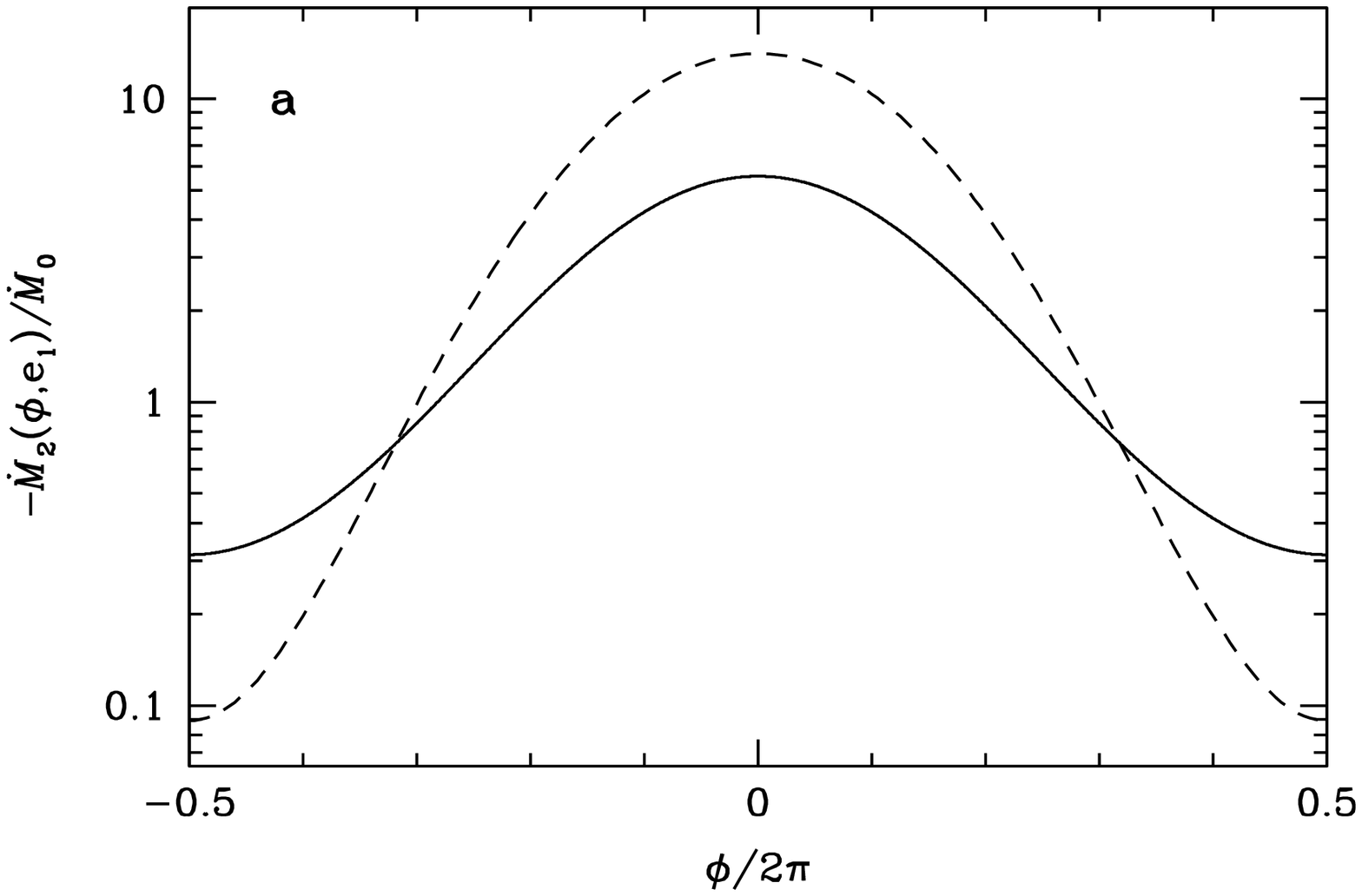}}
\centerline{\includegraphics[width=84mm]{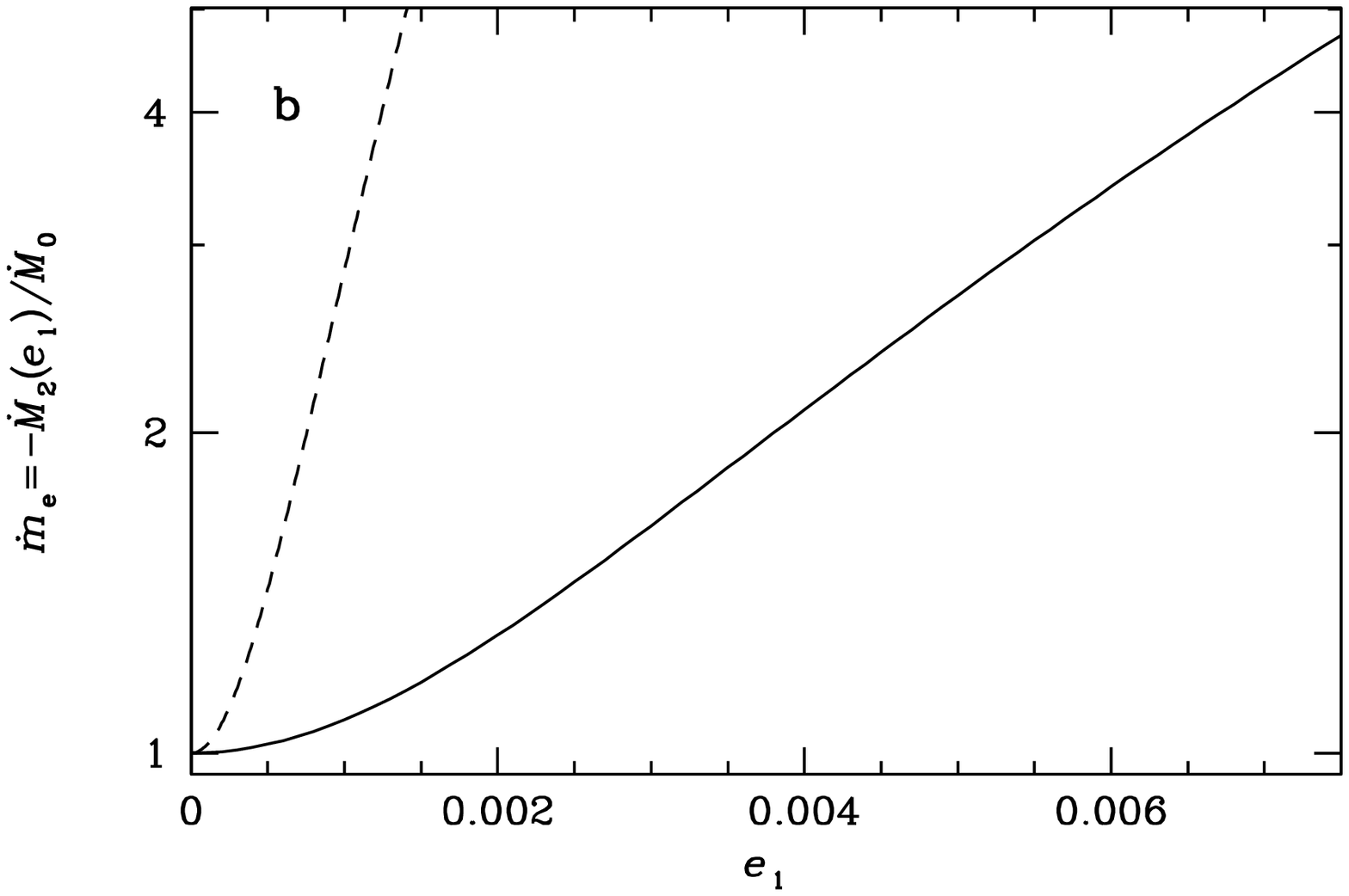}}
\caption{(a) The dashed and solid curves show the orbital phase dependence of the accretion rate for the model without irradiation at $e_1=8\times 10^{-4}$ and $d_0/R_0=1.64\times 10^{-3}$, equation (\ref{mdot_cold}), and with irradiation at $e_1=4\times 10^{-3}$ and $H_0/R_0=2.3\times 10^{-3}$, equation (\ref{mdot_exp_ratio}), respectively. The periastron corresponds to $\phi=0$. In both cases, $\dot m_e\simeq 2$, illustrating the orbital phase dependence of $\dot M_2$ at the respective $e_{\rm max}$ required to explain the superorbital cycle of \source. See equation (\ref{angular}) for ${\rm d}\phi/{\rm d}t$ (almost constant at $e_1\ll 1$). (b) The dashed and solid curves show the orbit-averaged accretion rate as function of $e_1$ for the two above models, respectively. 
\label{mdot_orbit} }
\end{figure}

For the assumed white dwarf parameters and the core temperature of $10^7$ K (for the model of Deloye \& Bildsten 2003; see above), we can calculate the intirinsic white dwarf luminosity [eq.\ (4.1.11) of Shapiro \& Teukolsky 1983]. For the metal abundance of $Z\sim 0.01$ of NGC 6624 given by Rich, Minniti \& Liebert (1993), $\mu=4/3$ for ionized He, and the (very approximate) Kramers' opacity of Schwarzschild (1958), we find that luminosity to be $\simeq 10^{-3}{\rm L}_{\sun}$. Using the above values, we then use the photon diffusion equation, hydrostatic equlibrium, and the equation of state and solve for the profiles of $T(d)$, $\propto d$, and $\rho(d)$, $\propto d^{3.25}$, of the nondegenerate white-dwar surface layer (Schwarzschild 1958; section 4.1 of Shapiro \& Teukolsky 1983). Then we find from equations (\ref{mdot}--\ref{area}) that $d_0\sim 1.64\times 10^{-3} R_0$. At this $d_0$, $T\simeq 1.9\times 10^5$ K and $\rho\simeq 7.0\times 10^{-7}$ g cm$^{-3}$. Note that $d_0$ is much less than the calculated thickness of the surface layer, $\simeq 0.01 R_0$, below which the electrons become partly degenerate at the interior temperature of $\simeq 10^7$ K. 

The accretion rate averaged over the orbital period is,
\begin{equation}
\dot M_2(e_1)={(1-e_1^2)^{3/2}\over 2\upi} \int_0^{2\upi} {\dot M_2(\phi, e_1)\over (1+ e_1 \cos \phi)^2} {\rm d}\phi.
\label{mdot_av}
\end{equation}
Note that since this rate is averaged over time, a factor of ${\rm d}t/{\rm d}\phi$, equation (\ref{angular}), appears in the integral over $\phi$. We will also use a dimensionless accretion rate, 
\begin{equation}
\dot m_e\equiv {-\dot M_2(e_1)\over \dot M_0}.
\label{mdot_dim_less}
\end{equation}
We find that a very low $e_{\rm max}\simeq 8\times 10^{-4}$ is sufficient to increase $-\dot M_2$ by a factor of 2. Fig.\ \ref{mdot_orbit}(a) shows $-\dot M_2(\phi, e_1)/\dot M_0$ of equation (\ref{mdot_cold}) at this value of $e_1$. Fig.\ \ref{mdot_orbit}(b) shows the $\dot m_e$ of this model. Both the strong changes of the accretion rate over an orbit and the fast increase of the orbit-averaged rate with $e_1$ are due to the very fast increase of the product $\rho c_{\rm s} d$ with $d$, $\propto d^{4.75}$, in the surface layer of the white dwarf (see above). We note that we have used here the structure of an isolated white dwarf, while the gravitational field close to $L_1$ is affected by the gravity of the neutron star. This would flatten the density profile and lead to a requirement of a somewhat higher $e_{\rm max}$. 

Moreover, the white dwarf is very close to the X-ray source, and thus it is irradiated. Irradiation could be avoided if the accretion disk was strongly flared and obscured the white dwarf from the X-ray source. However, the observed strong orbital modulation in the UV (Anderson et al.\ 1997) is well explained by reprocessing of X-rays by the white dwarf (Arons \& King 1993), which interpretation rules out strong flaring. Thus, we consider now the case with irradiation of the white dwarf.

The white dwarf subtends the solid angle $\simeq \upi R_2^2/a_1^2$, which, for our binary parameters, is $\simeq 7.3\times 10^{-3} 4\upi$. For the $\langle F_{\rm bol}\rangle$ of Z07 (see above) and an albedo of 0.5 (e.g., Anderson 1981; London, McCray \& Auer 1981), the white dwarf receives $\sim\! 2.2\times 10^{35}(D/7.6\,{\rm kpc})^2$ erg s$^{-1}$, i.e., several orders of magnitude more than its possible intrinsic luminosity. The absorbed flux per unit area at a mid-point of the white dwarf (at $\sim a_1-r_0/2$ from the neutron star) varies then in the range $F_{\rm irr}\simeq (0.9$--$1.8)\times 10^{16}(D/7.6\,{\rm kpc})^2$ erg cm$^{-2}$ s$^{-1}$. This variability is due to the X-ray luminosity undergoing the superorbital cycle with the amplitude of 2. The corresponding blackbody temperature changes from $T=T_0\simeq 1.1\times 10^5(D/7.6\,{\rm kpc})^{1/2}$ K to $2^{1/4}T_0$. Assuming $L\propto \dot m_e$, $T(e_1)=\dot m_e^{1/4} T_0$. In optically-thick regions, the structure of the atmosphere will be closely isothermal at this $T$ (e.g., Anderson 1981). Thus, the atmosphere density will decrease exponentially with the distance from the center of the white dwarf, $\rho \propto \exp(-d'/H)$, where $H$ is the scaleheight around the $L_1$ point, \begin{equation}
{H\over R_0} \simeq {c_{\rm s}^2 R_0\over GM_2 }
\simeq 2.3\times 10^{-3} \dot m_e^{1/4} {T_0\over 1.1\times 10^5\,{\rm K}},
\label{scaleheight}
\end{equation}
where we used our numerical values of the parameters for the second equality. Then, we use the results of Brown \& Boyle (1984), who derived ${\cal A}\simeq (2\upi)^{1/2} H R_0$ (see their eq.\ 4), which can be expressed as,
\begin{equation}
-\dot M_2(\phi, e_1) \simeq (2\upi)^{1/2} t_{\rm dyn}^2 c_{\rm s}^3 \rho_0 \exp(-d'/H),
\label{mdot_exp}
\end{equation}
where $t_{\rm dyn}=(R_0^3/ GM_2)^{1/2}$ ($\simeq 55$ s) is close to the dynamical time scale of the white dwarf [$(R_2^3/ GM_2)^{1/2}\simeq 30$ s], and $\rho_0$ is the atmosphere density at $L_1$ corresponding to $e_1=0$. Here we neglected the correction of Brown \& Boyle (1984) for the velocity of the $L_1$ point, since it is much slower than the sound speed in our case. Note that though the functional dependence $\propto c_{\rm s}^3 \rho_0$ is most likely accurate, the constant of the proportionality remains relatively uncertain (see Savonije 1983) as well as the argument of the exponent may be more complex than that used above, see Frank, King \& Raine (2002), pp.\ 352--353. The latter effect may increase the scaleheight around $L_1$ and thus lead to a requirement of a higher $e_{\rm max}$ than than that estimated by us below. To solve this problem accurately, hydrodynamical simulations (see, e.g., Reg{\"o}s, Bailey \& Mardling 2005) specific to \source\ (beyond the scope of this work) are needed. 

The $H$ and $c_{\rm s}$ depend on $e_1$ through $T(e_1)$, and we denote their values at $e_1=0$ as $H_0$ and $c_{{\rm s}0}$, respectively. The accretion rate relative to that at $e_1=0$ can then be written as,
\begin{equation}
{-\dot M_2(\phi,e_1)\over \dot M_0} \simeq \dot m_e^{3/8} \exp \left({e_1 R_0 \over H_0\dot m_e^{1/4}} {e_1+\cos\phi\over 1+e_1\cos\phi}\right),
\label{mdot_exp_ratio}
\end{equation}
This equation coupled with equations (\ref{mdot_av}--\ref{mdot_dim_less}), giving $\dot m_e$, can be easily solved iteratively.  

This solution, for a given $H_0/R_0$, can be used to calculate the eccentricity required for a given amplitude of $L$. Fig.\ \ref{mdot_orbit}(a) shows an example of the dependence of equation (\ref{mdot_exp_ratio}) for $e_1=0.004$ and $H_0/R_0=0.0023$, yielding $\dot m_e\simeq 2$ (i.e., $e_{\rm max}\simeq 0.004$). Fig.\ \ref{mdot_orbit}(b) shows $-\dot M_2(e_1)$ for the same $H_0/R_0$. These dependences are slower than those of the unirradiated case due to the slower (in the present case) increase of the density with the depth within the atmosphere as well its isothermality. 

In the irradiated case, the accretion rate enhancement due to the eccentricity depends primarily not on $e_1$ itself but on $e_1/(H_0/R_0)$. In particular, the eccentricity required to obtain the increase of the orbit-averaged accretion rate by 2 is $e_{\rm max}\sim 2H_0/R_0$. Also, $\dot m_e$ increases initially, at low values of $e_1$, very slowly. Thus, our assumption that $\dot M_0$ corresponds to $e_1=0$ (see above) introduces only a slight error as long as $e_{\rm max}\gg e_0$. We also note that the value of $e_{\rm max}$ does not depend on the (relatively uncertain) normalization of the $-\dot M_2$ dependence of equation (\ref{mdot_exp}).

From that normalization, we can determine $\rho_0$ as,
\begin{equation}
\rho_0={\dot M_0 \over (2\upi)^{1/2} t_{\rm dyn}^2 c_{{\rm s}0}^3}\simeq 2.0\times 10^{-6}{\rm g\,cm}^{-3}.
\label{rho}
\end{equation}
This corresponds to an electron density of $n_0\simeq 6.0\times 10^{17}$ cm$^{-3}$, and the Thomson optical depth above the $L_1$ of 1.0. The Rosseland absorption opacity at the above $\rho$, $T\simeq 1.1\times 10^5$ K and $Z=0.01$ is $\simeq 17\, {\rm g}^{-1}\, {\rm cm^2}$ (Iglesias \& Rogers 1996), i.e., $\sim 10^2$ more than the scattering opacity for He. Thus, the medium is completely optically thick, consistent with our assumption of the temperature equal to the blackbody one. The corresponding pressure is $\sim 10^7$ dyn cm$^{-2}$, which can be calculated to be $\sim\! 10^2$ times more than the critical pressure below which a transition to an optically thin regime begins (London et al.\ 1981). Note that the $L_1$ region may be partly shadowed by the accretion disc, which would decrease the temperature and move the $L_1$ region even more into the optically-thick regime. We also check that the time scale at which the atmosphere locally responds to the irradiation, $\sim 2n_0 k T_0 H_0/F_{\rm irr}\simeq 15$ ms, is much shorter than any other time scale of interest.

On the other hand, the coefficient in equation (\ref{mdot_exp}) and other details of the flow through $L_1$ are relatively uncertain, and thus it is of interest to also consider the case with the $L_1$ region being optically thin. Above a transition zone (Anderson 1981; London et al.\ 1981), the atmosphere temperature becomes close to the Compton temperature of the surrounding radiation field (e.g., Kallman \& McCray 1982; Begelman, McKee \& Shields 1983),
\begin{equation}
T_{\rm C}= {\int h\nu J_\nu {\rm d}\nu \over 4k \int J_\nu {\rm d}\nu},
\label{TC}
\end{equation}
where $J_\nu$ is the mean intensity. For the sum of the X-ray spectrum of the source, its reflection from the star, and the blackbody emission from reprocessing in the underlying regions of the star, $T_{\rm C}$ will be $\sim 2/3$ of the Compton temperature for the X-ray spectrum alone. Based on the spectral fits of Z07, we have calculated the latter to range from (1.6--$1.8)\times 10^7$ K in the high-luminosity state to $\sim 10^8$ K in the low-luminosity state. Thus, $T_{\rm C}\sim 10^7$ K for the dominant high-luminosity state. As the Compton temperature depends only on the shape of the spectrum but not on its flux, $T\simeq T_{\rm C}$ is almost independent of $-\dot M_2$, except for the lowest $-\dot M_2$ corresponding to the hard low state, where $T_{\rm C}$ is several times higher. Neglecting the last effect (important only at the flux minima), equation (\ref{scaleheight}) yields $H/R_0\sim 0.2$. 

If this condition were dominant over the superorbital cycle, a rather high $e_{\rm max}\sim 0.4$ would be required to account for the superorbital variability. This would require changing the radius (and mass) of the white dwarf, and raise the issue of the stability of the mass transfer (beyond the scope of this work). However, we stress that according to our estimates the $L_1$ region of the irradiated star is unlikely to be in the optically thin regime.

On the other hand, we cannot rule out that the $L_1$ point is in the transition zone (Anderson 1981; London et al.\ 1981) between $T\simeq 10^5$ K and $10^7$ K, in which case detailed solutions of the atmosphere structure would be required to calculate $-\dot M_2(e_1)$, and the $e_{\rm max}$ would be somewhat higher than the value of 0.004 estimated above. On the other hand, the shadowing of the white dwarf by the disc would (as mentioned above) decrease the temperature of the $L_1$ region. These effects introduce some systematic uncertainties on the value of $e_{\rm max}$.

\section{Models of the triple system}
\label{triple}

Above, we have modelled the accretion rate variability in \source\ as due to variability of the binary eccentricity. Here, we model the required variability of the eccentricity as due to gravity of a third star in the system. The third star should have the semimajor axis ($a_2$; measured with respect to the center of mass of the inner binary) satisfying $a_2/a_1\gg 1$ in order not to perturb the binary motion on time scales in the range between $P_1$ and $P_3$, which perturbations have not been observed. Thus, this model is of a hierachical triple system. We then attempt to reproduce the observed 171-d period and its relative coherence (Section \ref{data}) as well as the eccentricity amplitude necessary to reproduce the observed amplitude of the superorbital variability as inferred from our $L_1$-flow calculations (Section \ref{flow}).

We begin with a brief review of the main features of secular effects in hierarchical triple systems. It has been known (e.g., Kozai 1962) that a third outer body can induce quasi-periodic oscillations of the inner eccentricity with a long quasiperiod of
\begin{equation}
P_3= K {P_2^2\over P_1},
\label{period}
\end{equation}
where $P_2$ (the outer orbital period) is given by 
\begin{equation}
P_2^2={4\upi^2 a_2^3\over GM},
\label{P2}
\end{equation}
$M=M_1+M_2+M_3$, and $K$ (often $\sim$1) depends on the system parameters. There are two main regimes. At an initial mutual inclination, $i_0$, above a critical angle, $i_{\rm c}\simeq 40\degr$, the amplitude of the inner eccentricity is large and $K\sim 1$. The evolution can be well described by taking into account only the lowest-order perturbative term in the system Hamiltonian expanded in powers of $a_1/a_2$, namely the quadrupole term, $\propto (a_1/a_2)^2$. Then, the mutual inclination, $i$, is anticorrelated with the inner eccentricity, $e_1$, with $(1-e_1^2)\cos^2 i\simeq$ constant. In this approximation, both semimajor axes, the outer eccentricity, $e_2$, and the magnitude of the angular momentum of the outer binary are constant, and there is exchange of the angular momentum between the inner and outer binary. In a conservative system, also the total angular momentum and energy are constant. The maximum inner eccentricity is rougly $e_{\rm max}\simeq [1-(5/3)\cos^2 i_0]^{1/2}$ (Innanen et al.\ 1997; Holman, Touma \& Tremaine 1997) above the critical angle (and thus $e_{\rm max}\simeq 1$ for $i_0\simeq 90\degr$). 

Below $i_{\rm c}$, $e_{\rm max}\ll 1$, and the quadrupole approximation becomes insufficient. In particular, for coplanar orbits, $i_0=0$, the quadrupole term in the Hamiltonian becomes null, and the first non-zero term is octupole, $\propto (a_1/a_2)^3$. In the octupole approximation, only the semimajor axes are constant (apart from the total energy and angular momentum). In this regime, the inner eccentricity also varies quasi-periodically, but at a period longer than that of the quadrupole one. In the intermediate regime where both terms are important, time dependencies of the inner eccentricity show both periodicities (Krymolowski \& Mazeh 1999). Also, at $e_{\rm max}\ll 1$ in general, $e_{\rm max}$ decreases with the increasing $a_2/a_1$, increases with the increasing the initial outer eccentricity, $e_{2,0}$, and it depends very weakly on either $M_3/M_{12}$ or $M_2/M_1$ (F00).

We use a numerical model calculating the time evolution of an isolated hierarchical triple of point masses, using secular perturbation theory up to the octupole terms (e.g., F00; Miller \& Hamilton 2002; Blaes et al.\ 2002; W03). We neglect possible effect of the tidal deformation of the white dwarf on the gravitational force exerted on the inner binary, see, e.g., S\"oderhjelm (1984), Eggleton \& Kiseleva-Eggleton (2001). Also, the Hamiltonian is averaged over the periods of both the inner and outer binaries, and thus any possible short-time scale changes (e.g., Bailyn 1987; Georgakarakos 2002) are averaged out. Our calculations are Newtonian apart from the general relativistic (GR) periastron precession in the inner binary. Its first-order post-Newtonian period, $P_{\rm PN}$, is given by (e.g., Weinberg 1973),
\begin{equation}
P_{\rm PN}={P_1 a_1 c^2 (1-e_1^2)\over 3GM_{12}}= {P_1^{5/3} c^2(1-e_1^2)\over 3(2\upi GM_{12})^{2/3}}.
\label{pgr}
\end{equation}
We include this effect in the same way as in, e.g., W03.

The free parameters of the model are the masses of the neutron star, $M_1$, and of the third star, $M_3$, $a_2/a_1$, $i_0$, $e_{2,0}$, and $e_0$. We assume $0 \la e_0\ll e_{\rm max} \simeq 0.004$ (Section \ref{flow}). Then, we know $P_1$ and $P_3$, and assume $M_2=0.07\msun$ (Section \ref{flow}).

Given the relatively large number of free parameters and complex dependencies between them and the resulting period and the amplitude of the eccentricity modulation, it has proven difficult to constrain the parameter space numerically. Thus, we have obtained some approximate analytical constraints. Given the presence of only one, well-defined, long-term periodicity in the system (Section \ref{data}; W06), evolution of the triple system can be dominated by either the quadrupole or the octupole term, but not by both, which would have given rise to two periodicities (as noted above). Given the considerably greater simplicity of the quadrupole equations, we consider them for our analytical estimates (though we do include the octupole term in numerical calculations). Given our requirement of $e_{\rm max}\ll 1$, this implies values of $i_0$ close to $i_{\rm c}$. 

We first estimate the maximum eccentricity. We use the conservation of the total angular momentum and energy. The former gives us the mutual inclination, $i$, in terms of $i_0$ and $e_1$, e.g., eq.\ (3) in Miller \& Hamilton (2002). Then the (quadrupole) Hamiltonian gives the total energy. The minimum and maximum inner eccentricity corresponds to the inner periastron angle (measured from the intersection of the two orbits), $g_1$, of 0 and $\upi/2$, respectively. Equating the Hamiltonian at those two angles and using $i$ from the angular momentum conservation, gives, in the lowest (second) order of $e_1$, 
\begin{equation}
e_{\rm max}^2 \simeq  \frac{4+\theta_{\rm PN}+2\cos i_0/\beta}{10\cos^2i_0-6+\theta_{\rm PN}+2\cos i_0/\beta} e_0^2,
\label{e_max}
\end{equation}
where 
\begin{eqnarray}
\lefteqn{\theta_{\rm PN} \equiv \frac{8 G M_{12}^2 a_2^3 (1-e_{2,0}^2)^{3/2}} {c^2 M_3 a_1^4}, \label{tPN} }\\ 
\lefteqn{  \beta \equiv {M_{12}^{3/2} M_3 a_2^{1/2}(1-e_{2,0}^2)^{1/2}\over M^{1/2} M_1 M_2 a_1^{1/2}}.\label{beta} }
\end{eqnarray}
In order for a solution with $e_{\rm max} \gg e_0$ to exist, the denominator in equation (\ref{e_max}) should be nearly zero. Also, $\beta\gg 1$ at $a_2\gg a_1$, which we assume hereafter. Thus,
\begin{equation}
\cos^2 i_0 \simeq \cos^2 i_{\rm c} \simeq \frac{6-\theta_{\rm PN}}{10}.
\label{icrit}
\end{equation}
Note that this $i_{\rm c}$ is also the critical angle for the large $e_{\rm max}$ regime (see Blaes et al.\ 2002), and it becomes the Newtonian critical angle when $\theta_{\rm PN} =0$ (Kozai 1962). In the small $e_1$ limit, the initial mutual inclination angle should be approximately equal to but slightly {\it smaller\/} than the critical value, which is different from the requirement for high $e_{\rm max}$ regime. The value of $e_{\rm max}$ is determined by how close $i$ is to $i_{\rm c}$.  

For a solution of equation (\ref{icrit}) to exist, $\theta_{\rm PN} < 6$ is required. Then, we have from equation (\ref{tPN}), 
\begin{equation}
{a_2^3\over a_1^3}\la {3\over 4} {c^2 a_1 M_3\over 4 G M_{12}^2} (1-e_{2,0}^2)^{-3/2}.
\label{a_ratio}
\end{equation}
For $M_{12}\simeq 1.5$, the upper limit on $M_3$ of $0.5\msun$ (CG01) and $e_2^2\ll 1$, $a_2/a_1\la 25$. Note that the constraint of equation (\ref{a_ratio}) is the same as that derived for the high-$e_1$ case [except for the $(1-e_1^2)^{3/2}$ factor, Blaes et al.\ 2002]. It is related to the fact that the Newtonian regime corresponds to $P_{\rm PN}\gg P_3$. Otherwise, the GR periastron precession decreases the amplitude of the Kozai oscillations of $e_1$, and leads to its disappearing in opposite limit. This is because the effect of the third body is a coherent summation of the tidal perturbations over many orbital periods, and the GR precession partly destroys this coherence (e.g., W03). In particular, this leads to a change of $i_{\rm c}$, equation (\ref{icrit}), reducing the high-$e_{\rm max}$ regime. 

Another constraint on the parameter space comes from the observed $P_3\simeq 171$ d. The inner binary spends most of the time around $g_1 = \upi/2$, at which ${\rm d}g_1/{\rm d}t=0$. We thus write $g_1 = \upi/2+\delta$ with $\delta \ll 1$ and expand the (quadrupole) evolution equations for $e_1$ and $g_1$ [e.g., eqs.\ (16--17) in W03] to the first order in $\delta$ and in the small-$e_1$ approximation. We then use equation (\ref{e_max}) to express the results in terms of $e_{\rm max}/e_0$, divide ${\rm d}e_1/{\rm d}t$ by ${\rm d}g_1/{\rm d}t$, and integrate over $e_1$ from $e_0$ to $e_{\rm max}$. This yields the value of $K$ [equation (\ref{period})] of
\begin{equation}
K=\frac{2^{5/2} f}{3\upi} \frac{M}{M_3(4+\theta_{\rm PN})} (1- e_{2,0}^2)^{3/2} {e_{\rm max}\over e_0} \ln^{1\over 2} {e_{\rm max}\over e_0},
\label{P3_PN}
\end{equation}
where $f$ is a fudge factor to compensate for the approximation we used in the derivation. We found empirically that $f\simeq 1$ are within a factor of two, and in our example described below, $f \simeq 1.1$. Note also that the superorbital period is strongly dependent on $e_0$, on which the accretion rate depends very weakly (as long as $e_0\simeq 0$, Fig.\ \ref{mdot_orbit}b).

The following constraint on the total mass of the inner binary can be obtained using equations (\ref{period}--\ref{P2}), (\ref{tPN}) and (\ref{P3_PN}),
\begin{equation}
\frac{3(G\msun/c^3)^{2/3} (2\upi)^{5/3} P_3 e_0}{2^{1/2} P_1^{5/3} e_{\rm max} \ln^{1/2} (e_{\rm max}/e_0) } = \frac{f\theta_{\rm PN}}{4+\theta_{\rm PN}} \left (\frac{M_{12}}{\msun} \right )^{-2/3}\!, 
\label{P_PN}
\end{equation}
which (for the observed $P_1$ and $P_3$) can be solved for $\theta_{\rm PN}$ as,
\begin{equation}
\theta_{\rm PN} \simeq  {4\over 0.274 f \left(M_{12}/ \msun\right)^{-2/3}{e_{\rm max}\over e_0}\ln^{1\over 2} {e_{\rm max}\over e_0}-1}.
\label{t_solution}
\end{equation}
The constraint of $\theta_{\rm PN} < 6$ then yields a relation between $M_{12}$ and $e_{\rm max}/ e_0$,
\begin{equation}
{M_{12}\over \msun} \la 0.067 \left( f e_{\rm max}\over e_0\right)^{3/2} \ln^{3/4} {e_{\rm max}\over e_0}.
\label{mupper}
\end{equation}
At $f\simeq 1.1$, it allows $M_{12}\ga 1.5\msun$ provided $e_{\rm max}/e_0\ga 5.5$. Since the required relative amplitude of $\dot M_2$ of $\simeq 2$ can be achieved at any $e_{\rm max}/e_0\ga 3$ (see the solid curve in Fig.\ \ref{mdot_orbit}b), this is only a very weak constraint, allowing practically any of the theoretically possible masses of the neutron star at modest values of $e_{\rm max}/e_0$.

Equation (\ref{tPN}) at $\theta_{\rm PN} < 6$ also yields a constraint on $M_3$. If we assume $M_{12}\simeq 1.5\msun$ and that the system is hierarchical, $a_2/a_1\ga 5$, we obtain $M_3\ga 0.004\msun$ [see also equation (\ref{a_ratio})]. An independent relation follows from equations (\ref{period}--\ref{P2}) and (\ref{P3_PN}),
\begin{equation}
{M_3 \over (1-e_{2,0}^2)^{3/2} M_{12}}=  {2^{5/2}a_2^3 P_1 e_{\rm max}\over 3\upi a_1^3 (4+\theta_{\rm PN}) P_3 e_0}\ln^{1\over 2} {e_{\rm max}\over e_0},
\label{M3}
\end{equation}
which yields the same constraint of $M_3\ga 0.004\msun$ at $e_{\rm max}/e_0\ga 7$. Thus, even a very low-mass third body can still induce the required eccentricity oscillations in the inner binary. 

We have not studied analytically constraints in the octupole-dominated regime (which regime has been considered, e.g., by Rasio 1994, 1995; F00; Georgakarakos 2002; Lee \& Peale 2003). The octupole-dominated regime may likely yield solutions with $i_0\sim 0$ (as compared to $i_0\simeq i_{\rm c}$ in the quadrupole regime). Still, if such solutions exist, our conclusion of only very weak constraints on the masses of the system components would remain unaffected. On the other hand, an important difference between solutions in the two regimes is the variability of the mutual inclination. In the quadrupole low-$e_1$ regime, the maximum change, $\Delta i= i-i_0$, is given by
\begin{equation}
\Delta i\simeq -{e_{\rm max}^2-e_0^2\over 2}\cot i_0,
\label{delta_i}
\end{equation}
which implies only a very small change of $i$ over the course of the superorbital cycle, e.g., about $-2.3''$ at $i_0=40\degr$, $e_{\rm max}=0.004$. On the other hand, numerical results of F00 (fig.\ 8) show $\Delta i\sim \pm 15\degr$ in their example for the octupole-regime, low-$e_1$ oscillations, for which $i_0=15\degr$, $e_{\rm max}\simeq 0.001$. Since the angular momentum is dominated by the outer binary, $i$ is approximately equal to the inclination with respect to the constant axis of the total angular momentum, which, in turn, is related to the value of (in principle observable) the inclination of the inner orbit to the line of sight.

For assumed values of $M_1$, $M_2$, $M_3$, $e_0$, $e_{\rm max}$, and with the observed $P_1$ and $P_3$, the above equations can be used to find $a_2$ and $i_0$. These appproximate analytical solutions can be then fine-tuned numerically. We first present an example yielding $P_3\simeq 171$ d within $\sim$1 per cent and $e_{\rm max}\simeq 0.004$, corresponding to our model of the accretion with irradiation (Section \ref{flow}). Its parameters are $M_1=1.29\msun$, $M_2=0.07\msun$, $e_0=10^{-4}$, $M_3= 0.5\msun$, $a_2/a_1=8.66$ (yielding $P_2\simeq 0.17$ d), $i_0=40.96\degr$, $e_{2,0}=10^{-4}$, corresponding to $\theta_{\rm PN} \simeq 0.22$, $\beta\simeq 19$, and $K\simeq 41$. The initial values of the periaxis angles, $g_1$ and $g_2$, are set to zero. Note that $e_{\rm max}/e_0$ is very sensitive to $i_0$, see equation (\ref{e_max}). 

\begin{figure}
\centerline{\includegraphics[width=84mm]{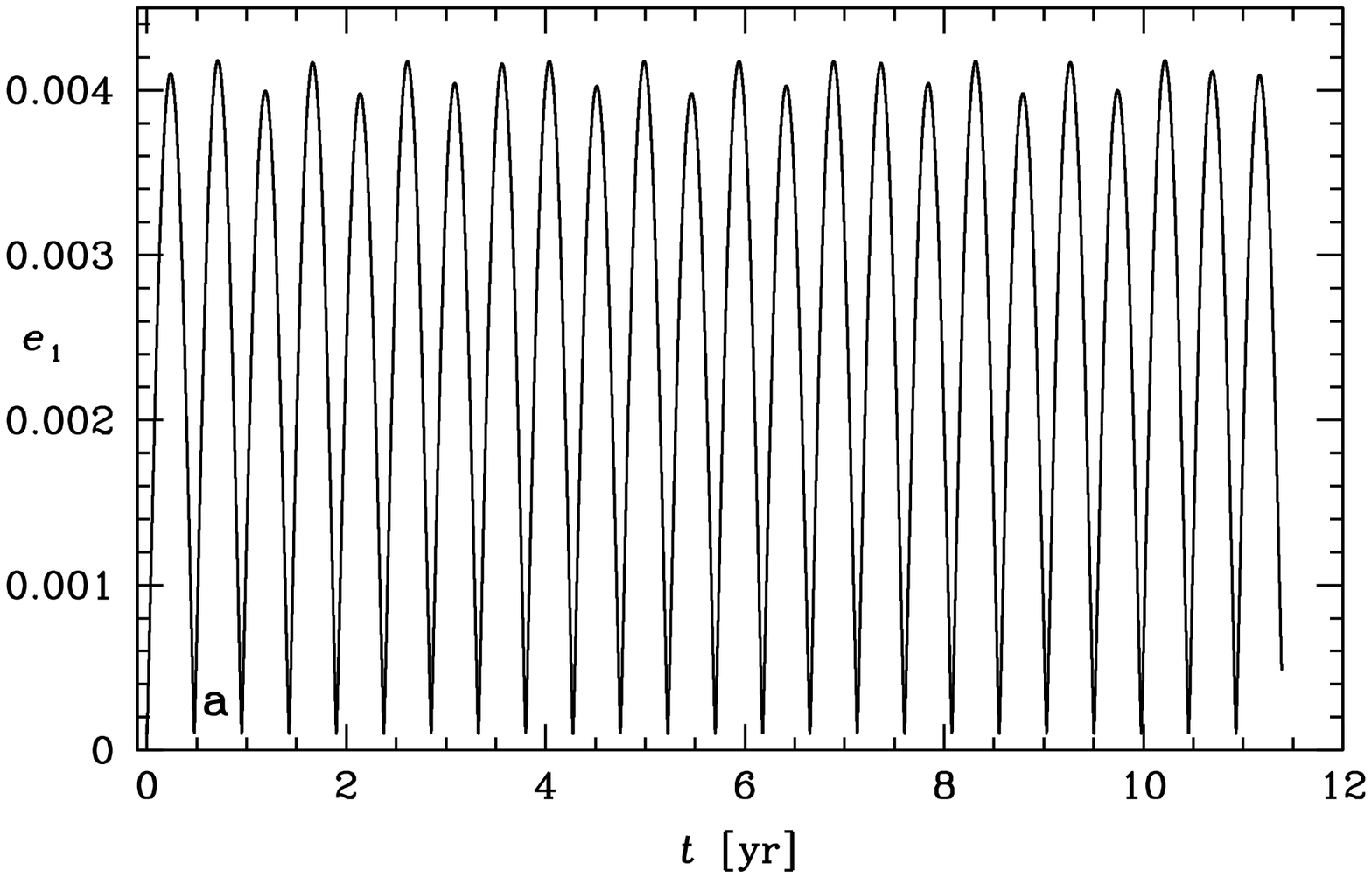}}
\centerline{\includegraphics[width=84mm]{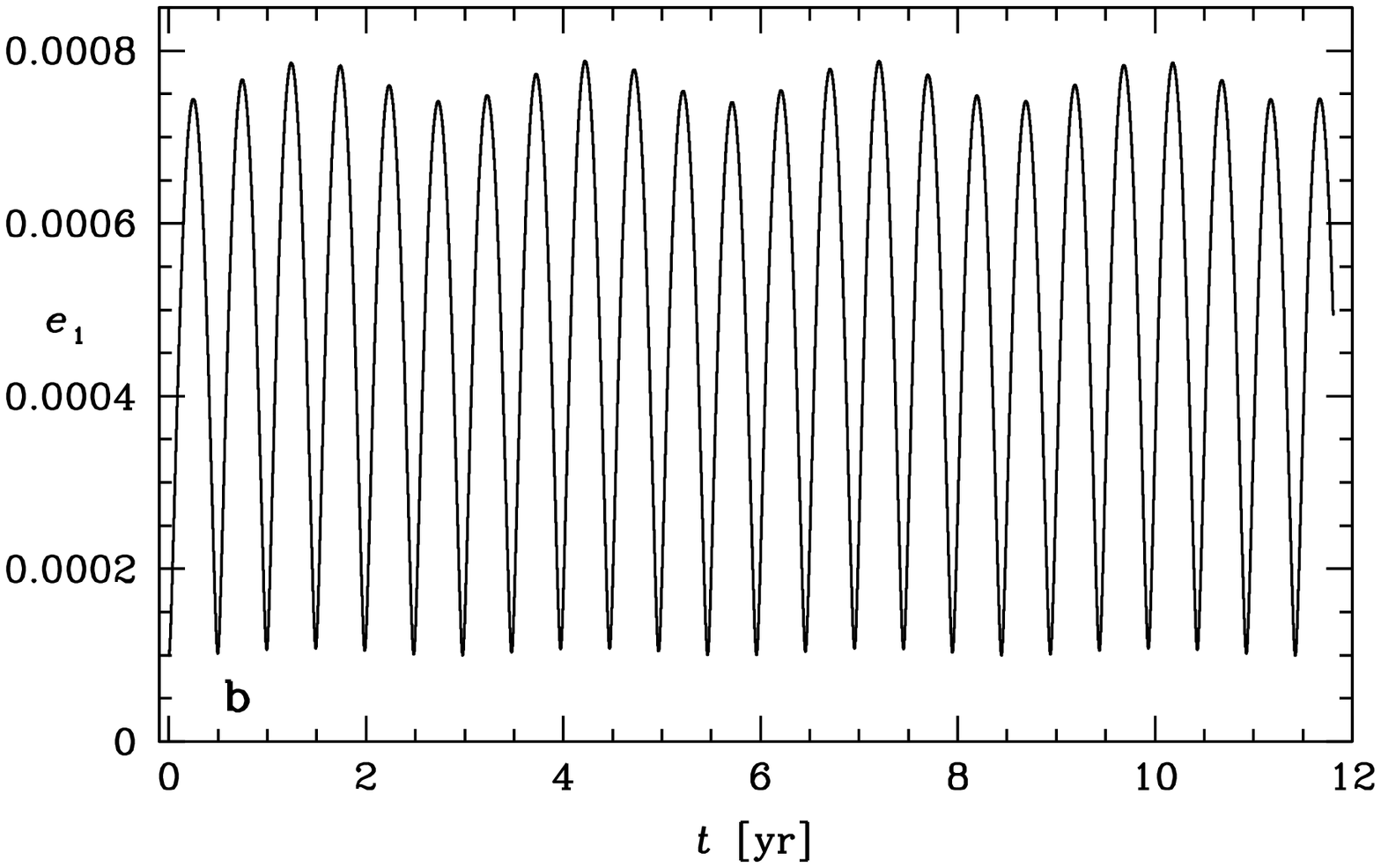}}
\caption{The time evolution of the inner eccentricity, $e_1$, for our best model (a) with irradiation of the white dwarf and (b) without irradiation.
\label{f_triple} }
\end{figure}

\begin{figure}
\centerline{\includegraphics[width=70mm]{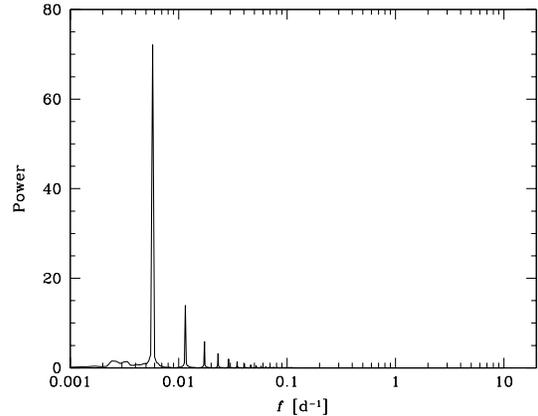}}
\caption{The Fourier power spectrum of the eccentricity dependence for the case with irradiation, Fig.\ \ref{f_triple}(a).
\label{f_power} }
\end{figure}

Time evolution of this model is shown in Fig.\ \ref{f_triple}(a). Fig.\ \ref{f_power} shows the Fourier power spectrum of $e_1$ at the sampling rate of 20 d$^{-1}$ (to approximate that of the \xte/ASM). The $P_3$ peak is unresolved at the resolution of the Fourier spectrum. This narrowness is similar to that observed, cf.\ fig.\ 19 of W06. In Fig.\ \ref{f_triple}(a), we also see a weak second quasi-periodicity, with the period slightly longer than twice of the main one. This is an effect of the octupole term in the evolution equations (see above). 

We then use our results of Section \ref{flow} connecting $e_1$ to $-\dot M_2$. We first apply equations (\ref{mdot_exp_ratio}) and (\ref{mdot_av}--\ref{mdot_dim_less}), yielding $-\dot M_2(e)$ with the flow parameters as in Fig.\ \ref{mdot_orbit} to convert $e_1(t)$ of Fig.\ \ref{f_triple}(a) to $-\dot M_2(t)$. We multiply the model period by 0.987 in order to obtain the exact agreement with the observed $P_3$. We first plot the resulting $-\dot M_2(t)$ (fitting only the normalization) by the solid curve in Fig.\ \ref{asm}. We see that the model fits well the overall superorbital variability. We also note that the exact Newtonian calculations of the evolution would increase the scatter in $e_1(t)$, i.e., leading to a less periodic behaviour (e.g., F00). This could explain the presence of some observed superorbital cycless with a significantly different duration that the average $P_3$ (see Fig.\ \ref{asm}).

We then take the second cycle in Fig.\ \ref{f_triple}(a) as representative and fit it to the average superorbital phase diagrams of Figs.\ \ref{phase}(a--b), with the second free parameter being the phase offset (with respect to the ephemeris of CG01), which we find $\phi_{\rm 3,0}/2\upi=-0.088$ and $-0.076$ for the ASM and PCA data, respectively. The fit results, shown by the solid curves in Fig.\ \ref{phase}(a--b), reproduce well the average phase diagrams.

Finally, we also find a triple solution corresponding to our model without irradiation of the white dwarf, which requires $e_{\rm max}\simeq 8\times 10^{-4}$ (Section \ref{flow}). Its parameters are identical to the previous model except for $a_2/a_1=16.95$ (yielding $P_2\simeq 0.47$ d) and $i_0=48.5\degr$, corresponding to $\theta_{\rm PN} \simeq 1.65$, $\beta\simeq 26.5$, and $K\simeq 4.6$. Time evolution of this model is shown in Fig.\ \ref{f_triple}(b). 

\section{Long-term evolution and other constraints}
\label{long-term}

Above, we have neglected the effects of emission of gravitational radiation (and of any other process of long-term loss of angular momentum) on the evolution of the inner binary. This is fully justified as we have considered evolution over the course of a decade, which is many orders of magnitude shorter than the time scale due to emission of gravitational waves. Thus, we can consider the long term evolution separately from the short-term one. 

The time scale for the loss of the angular momentum, $J_1$, from the inner system due to emission of gravitational radiation is given by (Peters 1964),
\begin{eqnarray}
\lefteqn{ \tau_{\rm g}^{-1}=-{\dot J_{\rm g}\over J_1}={32\over 5} \left(2\upi\over P_1\right)^{8/3} {G^{5/3}\over c^5} {M_1 M_2\over M_{12}^{1/3}} {1+(7/8)e_1^2\over (1-e_1^2)^{5/2}}
} \\
\lefteqn{\qquad
\simeq 1.23\times 10^{-6} {M_1 M_2\over \msun^{5/3} M_{12}^{1/3}} \left(P_1\over 1\,{\rm s}\right)^{-8/3} {\rm s}^{-1}, \quad e_1\ll 1,
\label{tau_g}
}
\end{eqnarray}
where
\begin{equation}
J_1=M_1 M_2 \left[G a_1(1-e_1^2)\over M_{12}\right]^{1/2}.
\label{J}
\end{equation}
This yields $\tau_{\rm g}\simeq 1.1\times 10^7$ yr for our system parameters. We see that the effect of $e_1>0$ on the rate of emission of gravitational radiation is negligible as long as $e_1^2\ll 1$. 

By logarithmically differentiating equation (\ref{J}), assuming conservative mass transfer ($\dot M_1=-\dot M_2$) and using equations (\ref{roche}--\ref{kepler}) and (\ref{rwd}), we can relate $\dot J_1$ to $\dot M_2$ and $\dot P_1$ (assuming $e_1^2\ll 1$),
\begin{equation}
-{\dot J_1\over J_1}={-\dot M_2\over M_2} \left({2\over 3}-{M_2\over M_1}\right) = {\dot P_1\over P_1}\left({2\over 3}-{M_2\over M_1}\right),
\label{evolution}
\end{equation}
where $\dot J_1$ includes all contributions to the inner angular momentum loss. Note that although the white dwarf expands as it loses mass, the accretion from $M_2$ is not self-sustaining, i.e., $\dot M_2=0$ if $\dot J=0$, as well as it can take place only for $M_2/M_1<2/3$. Also, $\dot P_1/P_1=-\dot M_2/M_2>0$. This is in conflict with the observations, showing $\dot P_1<0$ (see Section \ref{intro}). We note here that taking into account a possible mass loss from the system cannot resolve this discrepancy. If a fraction, $\delta$, of the mass lost by the secondary is ejected from the system, we derive from equations (\ref{roche}--\ref{kepler}) and (\ref{rwd}), 
\begin{equation}
{\dot P_1\over P_1}={-\dot M_2\over M_2}\left(1-\delta{M_2\over M_{12}}\right).
\label{pdot}
\end{equation}
Thus, although the mass loss reduces $\dot P_1$, it is still $>0$. Only in the absence of mass transfer, when the companion does {\it not\/} fill its Roche lobe, angular momentum losses (in particular, those due to gravitational radiation) lead to $\dot P_1/P_1=3\dot J/J<0$. A possible resolution of this conflict is the apparent observed $\dot P_1<0$ resulting from gravitational acceleration of the system in the cluster potential (CG01). 

Then, the likely evolutionary scenario of \source\ is still that of R87, as shown on their fig.\ 1. According to it, the period shortly after the onset of the mass transfer ($\sim\! 10^6$ yr ago) was $\sim$500 s. Then, $a_1$ was lower and $\theta_{\rm PN}$ higher [equation \ref{tPN})] than now. Consequently, $\cos^2 i_{\rm c}$ was less than the present value (i.e., the low-$e_{\rm max}$, octupole, regime was increased), see equation (\ref{icrit}). Thus, the eccentricity oscillations had in the past lower amplitudes than now, and the system was in the octupole-dominated regime. On the other hand, a future increase of $P_1$ may lead to $i>i_{\rm c}$ and moving the system to the high-$e_{\rm max}$ regime (provided $i_0\sim i_{\rm c}$ at present). 

On the other hand, we find that at the present moment of the source evolution, the GR periastron precession period, $P_{\rm PN}$, is very close, and may be equal exactly,  to the superorbital period, $P_3$. Namely,
\begin{equation}
{P_{\rm PN}\over 171\,{\rm d}}= (1-e_1^2)\left( P_1\over 685\,{\rm s}\right)^{5/3} \left(M_{12}\over 1.347\msun \right)^{-2/3}.
\label{pgr0}
\end{equation}
We consider this equality to be a very remarkable coincidence. We do not see any straightforward way in which the GR periastron precession could by itself (i.e., without the presence of a third star) affect the orbit-averaged accretion rate. Thus, the origin of the above equality can be either purely accidental or it may be due to some evolutionary processes not understood yet. We note that F00 find the presence of a resonance at $P_{\rm PN}\simeq P_3$ manifesting itself as a peak in $e_{\rm max}$ (fig.\ 14 in F00). However, it is not clear how that resonance could lead to the synchronization of $P_{\rm PN}\simeq P_3$.

On the other hand, our results and those of R87 do allow $P_{\rm PN}= P_3$ to be satisfied exactly. If this is the case and at $M_2\simeq (0.06$--$0.08) \msun$, $e_1\ll 1$ (Section \ref{flow}), the mass of the neutron star would be $M_1\simeq (1.27$--$1.29)\msun$. Remarkably, modelling of an X-ray burst of \source\ by Shaposhnikov \& Titarchuk (2004) gives $M_1 = 1.29^{+0.19}_{-0.06} \msun$. Also, Grindlay et al.\ (1984) have measured the total masses of low-mass X-ray binaries in globular clusters (including \source) and concluded that the initial neutron-star masses appear substantially less than $1.4\msun$ on average. We also note that some low neutron-star masses have recently been measured with high precision, e.g., $1.250^{+0.005}_{-0.005}\msun$ for PSR J0737--3039B (Lyne et al.\ 2004), and $1.18^{+0.03}_{-0.02} \msun$ for the neutron-star companion of PSR J1756--2251 (Faulkner et al.\ 2005). Given those results, $M_1=1.28\msun$ in \source\ appears entirely plausible.

We also note that although Arons \& King (1993) proposed that $M_1\sim 2\msun$ if the initial mass of the white dwarf were $\sim 0.6\msun$ and $\sim 0.5\msun$ has been accreted, the calculations of R87 imply a much lower initial mass and a short mass-transfer time interval, with only $\ll 0.1\msun$ accreted. Then, Zhang et al.\ (1998) proposed that $M_1\simeq 2.2\msun$ if the frequency of the upper kHz QPO at its apparent saturation at 1060 Hz were equal to the Keplerian frequency at the last stable orbit. However, van der Klis (2000) shows that there is no saturation in a more extensive data set. Also, current theoretical models of kHz QPOs do not postulate that frequency identification, and use it only to provide an upper mass limit (e.g., van der Klis 2000). 

The presence of the third star may affect the system evolution only if it remains stable and is not disrupted by encounters with stars in the cluster. 
The triple system itself is stable if $a_2/a_1\ga 2.8 (1+M_3/M_{12})^{2/5}$ (Mardling \& Aarseth 2001), which is clearly satisfied in our case, with $a_2/a_1\sim 10$, $M_{12}\sim 1.5\msun$ (Section \ref{triple}), and $M_3\la 0.5\msun$ (CG01). Then, the encounter time scale has been estimated by Miller \& Hamilton (2002) in the limit of domination of gravitational focusing, and we write it as
\begin{equation}
\tau_{\rm enc} \simeq 4\times 10^8 {4\times 10^5\,{\rm pc}^{-3}\over n_{\rm GC}} {10^{11}\,{\rm cm}\over a_2} {2\msun\over M}\, {\rm yr}.
\label{tau_enc}
\end{equation}
Here, $n_{\rm GC}$ is the number density of stars, which we assume to be numerically equal to the mass density of $\simeq 4\times 10^5\msun$ pc$^{-3}$ in the core of NGC 6624 (Ivanova et al.\ 2005), and $a_2\sim 10^{11}$ cm, $M\sim\! 2\msun$ (Section \ref{triple}). Thus, $\tau_{\rm enc}$ is much longer than the gravitational-radiation time scale, equation (\ref{tau_g}), but still shorter than an estimated cooling time of the white dwarf in \source\ of $\sim\! 10^9$ yr (the appendix of R87). Thus, an encounter could have formed the triple system during the lifetime of \source, but it is stable on the current evolutionary time scale.  

We also note that the angular momentum will also be lost from the system due to tidal dissipation within the secondary at $e_1>0$. This mechanism usually leads to circularization (and synchronization) of close binaries (e.g., Zahn 1977), but in our case the influence of the tertiary will force $\langle e_1\rangle>0$ and continuous dissipation (Mazeh \& Shaham 1979). The time scale for this process, $\tau_{\rm t}=-J_1/\dot J_{\rm t}$, is related to the circularization time scale. This, unfortunately, remains very uncertain for stars without convective envelopes, in particular white dwarfs (e.g., Zahn 1977), and it may be very long. Here, we consider the turbulent dissipation model of Press, Wiita \& Smarr (1975), which probably gives the the lower limit to the actual dissipation time scale in a white dwarf (Zahn 1977). Using the time scale of Press et al.\ (1975) and the approach of Mazeh \& Shaham (1979), we obtain,
\begin{equation}
\tau_{\rm t}= {125\upi\over 121} {(1-e_1^2)^{11/2}\over e_1^3} \left(a_1 \over R_2\right)^{11} {R_{\rm T}\over K_\mu} {M_2^3\over M_1^2 M_{12}} P_1,
\end{equation}
where $R_{\rm T}$ is the effective Reynolds number, which we assume $=20$ following Press et al.\ (1975), and $K_\mu$ is a dimensionless factor of the mean turbulent viscosity,
\begin{equation}
K_\mu={14\over M_2 R_2^{11}} \int_0^{R_2} \rho(r) r^{13} {\rm d}r,
\end{equation}
where $\rho(r)$ is the stellar density. We have calculated $K_\mu\simeq 0.0257$ for a polytrope with $n=3/2$, appropriate for a low-mass white dwarf. For our system parameters, we find then $\tau_{\rm t}\simeq 2300/\langle e^3\rangle$ yr (where the average is over the superorbital cycle), which for $\langle e_1^3\rangle^{1/3}= 0.003$ becomes $\sim 10^{11}$ yr. Thus, even if the dissipation is as fast as envisaged by Press et al.\ (1975), the process is completely negligible compared to the gravitational radiation. However, $\tau_{\rm t}<\tau_{\rm g}$ for $\langle e_1^3\rangle^{1/3}\ga 0.06$.

On the other hand, the soft X-ray absorption towards the source is consistent with being caused entirely by the interstellar medium, with no evidence for any local component (Futamoto et al.\ 2004). This appears to rule out strong outflows from either the irradiated disc, the irradiated surface of the white dwarf, or, in particular, mass loss through the $L_2$ point. At the mass ratio of the binary, the absence of the $L_2$ mass loss by the Roche-lobe filling star
imposes the constraint of $e_{\rm max}\la 0.07$, as calculated by R{\"e}gos, Bailey \& Mardling (2005). This rules out the above condition of $\langle e_1^3\rangle^{1/3}\ga 0.06$. Thus, presently, tidal dissipation appears not to be the dominant channel of angular momentum loss.

\section{Conclusions}

We have obtained the following main results.

1. Using the \xte\/ data, we quantify the average amplitude of the superorbital variability of the source. We find that amplitude in the folded and averaged light curves to be by about a factor of two. 

2. We consider the model in which the superorbital variability is due to a third star inducing cyclic variations (Kozai 1962) of the eccentricity of the binary (as proposed by CG01). We have studied the dependence of the rate of the Roche-lobe overflow on the eccentricity taking into account the strong irradiation of the white dwarf by the X-ray source (Fig.\ \ref{mdot_orbit}b). We find the amplitude of the eccentricity required to account for the variability of the accretion rate by the factor of two is $e_{\rm max}\simeq 0.004$ (assuming the minimum of the eccentricity is close to null). However, systematic uncertainties of this model may somewhat change the actual value of $e_{\rm max}$.

3. We reproduce the above maximum eccentricity of $\simeq 0.004$ and the observed superorbital period of 171 d in a detailed model of a hierarchical triple system. Our calculations are Newtonian except for inclusion of the GR periastron precession. We then convolve the obtained eccentricity light curve with our theoretical dependence of the accretion rate on the eccentricity and obtain a theoretical luminosity light curve. This theoretical light curve is compared to and found to be in a good agreement with the observed light curves from the ASM and PCA detectors, see Figs.\ \ref{asm}--\ref{phase}. 

4. We also study the parameter space allowed by the observational data, the $e_{\rm max}\simeq 0.004$ constraint, and $M_3\la 0.5\msun$ (CG01). We obtain analytical solutions for the low-eccentricity Kozai oscillations in the regime dominated by the quadrupole terms of the evolution equations. We find the ratio of the semimajor axes to be $\la 25$, which follows from the constraint that the GR periastron precession does not quench the Kozai eccentricity modulation. The masses of the system components are only weakly constrained. In particular, the canonical neutron-star mass of $1.4\msun$ is allowed in this model. The mass of the third body is constrainted as $0.004\la M_3/\msun \la 0.5$.

5. We find that the period of the binary GR periastron precession is approximately equal (and it is allowed to be exactly equal) to the observed superorbital period (which equals the period of the Kozai eccentricity oscillations in our model). We find this to be a remarkable and puzzling example of synchronization in a physical system. 

6. We obtain some other constraints on the system. The binary eccentricity has to be $\la 0.07$ from the apparent absence of the mass loss by the Roche-lobe filling white dwarf through the $L_2$ point. The angular momentum loss due to tidal dissipation in the white dwarf is found to be negligible compared to the loss due to emission of gravitational radiation. Also, we find the theoretical mass transfer rate due to the angular momentum loss via gravitational radiation is in complete agreement with that corresponding to the average bolometric flux from this source (as measured by Z07).

\section*{ACKNOWLEDGMENTS}

We thank M. Abramowicz, W. Dziembowski, M. Gilfanov, P. Haensel, J. Miko{\l}ajewska, C. Miller, A. Neronov, and A. Pamyatnykh for valuable discussions. We also thank the referee for constructive and valuable comments. This research has been supported in part by the Polish grants 1P03D01827, 1P03D01128 and 4T12E04727. LW acknowledges support by the Alexander von Humboldt Foundation's Sofja Kovalevskaja Programme (funded by the German Federal Ministry of Education and Research). MG acknowledges support through a PPARC PDRF. We also acknowledge the use of data obtained through the HEASARC online service provided by NASA/GSFC.

\label{lastpage}
\end{document}